\documentclass[twocolumn,showpacs,preprintnumbers,amsmath,amssymb,prl]{revtex4-1}
\usepackage{amssymb}
\usepackage{epsfig,amsmath,graphics}
\usepackage{epstopdf}
\usepackage{graphicx}

\renewcommand{\v}[1]{\ensuremath{\mathbf{#1}}}

\newcommand{\vertical}{{\textsc{\tiny $V$}}}
\newcommand{\horizontal}{{\textsc{\tiny $H$}}}
\newcommand{\coriolis}{{\textsc{\tiny $C$}}}
\newcommand{\earth}{{\textsc{\tiny $E$}}}

\def\be{\begin{equation}}
\def\ee{\end{equation}}

\DeclareMathOperator{\arccot}{arccot}

\begin{document}

\title{Enhanced atom interferometer readout through the application of phase shear}

\author{Alex Sugarbaker, Susannah M. Dickerson, Jason M. Hogan, David M. S. Johnson, and Mark A. Kasevich}
\email{kasevich@stanford.edu}
\affiliation{Department of Physics, Stanford University, Stanford, California 94305}

\date{\today}

\begin{abstract}
We present a method for determining the phase and contrast of a single shot of an atom interferometer.  The application of a phase shear across the atom ensemble yields a spatially varying fringe pattern at each output port, which can be imaged directly.  This method is broadly relevant to atom interferometric precision measurement, as we demonstrate in a $10\,\text{m}$ $^{87}\text{Rb}$ atomic fountain by implementing an atom interferometric gyrocompass with $10~\text{millidegree}$ precision.
\end{abstract}


\maketitle

Light-pulse atom interferometers use short optical pulses to split, redirect, and interfere freely-falling atoms \cite{Berman1997}. They have proven widely useful for precision metrology. Atom interferometers have been employed in measurements of the gravitational \cite{Fixler2007,Lamporesi2008} and fine-structure \cite{Bouchendira2011} constants, in on-going laboratory tests of the equivalence principal \cite{Hogan2009} and general relativity \cite{Dimopoulos2007, Hohensee2011}, and have been proposed for use in gravitational wave detection \cite{Dimopoulos2008b,Graham2013}. They have also enabled the realization of high performance gyroscopes \cite{Gustavson1997}, accelerometers \cite{Geiger2011}, gravimeters \cite{Peters2001}, and gravity gradiometers \cite{McGuirk2002}.

Current-generation light-pulse atom interferometers determine phase shifts by recording atomic transition probabilities \cite{Berman1997}.  These are inferred from the populations of the two atomic states that comprise the interferometer output ports. Due to experimental imperfections, interference contrast is not perfect -- even at the extremes, the dark port does not have perfect extinction.  This results in the need to independently characterize contrast prior to inferring phase.  Typically, this is done with a sequence of multiple shots with different phases, such that the population ratio is scanned through the contrast envelope \cite{Kasevich1992}. Such an experimental protocol relies on the stability of the contrast envelope. In many cases, the contrast varies from shot to shot, introducing additional noise and bias in the phase extraction process.

\begin{figure}
\begin{center}
\includegraphics[width=\columnwidth]{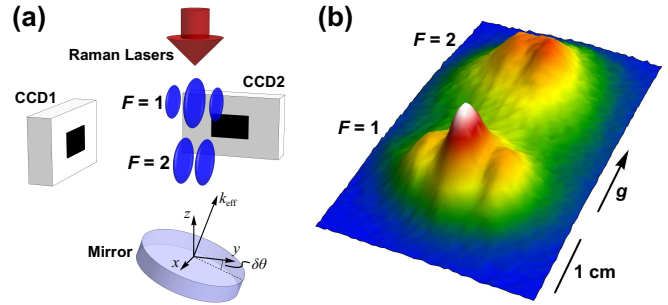}
\caption{\label{Fig:Apparatus}(a) Schematic diagram of the apparatus, showing beam-tilt phase shear readout.  Atoms are cooled and launched upward into an interferometer region, not shown.  Once they fall back to the bottom, the wavepackets are overlapped and an interference pattern (blue fringes) is imaged by two perpendicular cameras (CCD1,2).  An additional optical pulse is used to separate the two output ports ($F=1$ and $F=2$) by pushing the $F=2$ atoms downwards.  All atom optics pulses are performed by lasers incident from above and retroreflected off of a piezo-actuated mirror.  Tilting this mirror by an angle $\delta \theta$ for the third atom optics pulse yields a phase shear.  (b) A fluorescence image of the atomic density distribution taken with CCD2 after interference.  Spatial fringes result from a third-pulse tilt $\delta \theta = 60~\mu\text{rad}$ about the $x$-axis.  The pushed $F=2$ atoms are heated, yielding reduced apparent contrast, and we ignore the $F=2$ output port in subsequent analysis.}
\end{center}
\end{figure}

We present a broadly applicable technique that is capable of resolving interference phase on a single experimental shot.  This is accomplished through the introduction of a phase shear across the spatial extent of the detected atom ensemble.  The shear is manifest in a spatial variation of the atomic transition probability, which, under appropriate conditions, can be directly observed in an image of the cloud [Fig.~\ref{Fig:Apparatus}(b)].  Using this phase shear readout (PSR), it is no longer necessary to vary the phase over many shots to determine the contrast envelope.  Instead, the contrast of each shot can be inferred from the depth of modulation of the spatial fringe pattern on the atom ensemble.  The interferometer phase is directly determined from the phase of the spatial fringe.

The analysis of PSR fringes reveals rich details about atom interferometer phase shifts and systematic effects, much as the analysis of a spatially varying optical interference pattern yields information about the optical system and its aberrations.  The intentional application of a phase shear is analogous to the use of an optical shear plate, where a large applied phase shear highlights small phase variations across a laser beam.

In this work, we show that beam pointing can be used to introduce shear in a way that is broadly applicable to existing interferometer configurations.  In particular, this method does not require Bose-Einstein condensed or ultra-cold atomic sources.  We demonstrate the power of PSR by implementing a precise atom interferometer gyrocompass.  We also show how laser beam pointing and atom-optics pulse timing asymmetry can be combined to provide arbitrary control over the phase shear axis in the limit where the atoms expand from an effective point source.

The apparatus and methods are similar to those of our previous work \cite{Dickerson2013}. Using evaporative cooling followed by a magnetic lens, we obtain a cloud of $4 \times 10^6$ $^{87}\text{Rb}$ atoms with a radius of $200~\mu\text{m}$ and a temperature of $50~\text{nK}$.  These atoms are prepared in the magnetically insensitive $\left|F=2,\: m_F=0\right\rangle$ state, and then launched vertically into an $8.7~\text{m}$ vacuum tube with a chirped optical lattice. The atoms fall back to the bottom after $2.6~\text{s}$, and we then use a vertical fluorescence beam to image them onto two perpendicular CCD cameras (Fig.~\ref{Fig:Apparatus}).

While the atoms are in free-fall in a magnetically shielded region \cite{Dickerson2012}, we perform light-pulse atom interferometry with a $\pi/2-\pi-\pi/2$ acceleration-sensitive configuration with an interferometer duration of $2\,T=2.3~\text{s}$.  The atom optics pulses are applied along the vertical axis using two-photon Raman transitions between the $\left|F=2,\: m_F=0\right\rangle$ and $\left|F=1,\: m_F=0\right\rangle$ hyperfine ground states (the lasers are detuned $1.0~\text{GHz}$ blue of the $\left|F=2\right\rangle \rightarrow \left|F'=3\right\rangle$ transition of the D$_2$ line).  The atom optics light is delivered from above and retroreflected off of an in-vacuum piezo-actuated tip-tilt mirror.

The effective wavevector $\textbf{k}_\text{eff}$ of the Raman transitions is determined by the pointing direction of the retroreflection mirror \cite{Hogan2009}, which is set by the piezo stage for each atom-optics pulse with $1~\text{nrad}$ precision.  We compensate for phase shifts arising from the rotation of the Earth by applying additional tilts to each of the three pulses, as described in Refs.~\cite{Hogan2009, Dickerson2013}, but the mirror angle can also be used to induce shear for PSR.

To generate a controlled phase shear, we tilt the mirror for the final $\pi/2$ pulse by an angle $\delta \theta$ with respect to the initial two pulses (in addition to the tilts needed for rotation compensation).  In the semi-classical limit, the phase shift for a three-pulse interferometer is $\Delta \Phi = \textbf{k}_1 \cdot \textbf{x}_1 - 2 \textbf{k}_2 \cdot \textbf{x}_2 + \textbf{k}_3 \cdot \textbf{x}_3$, where $\textbf{k}_i \equiv \textbf{k}_{\text{eff},i}$ is the effective propagation vector at the time of the \textit{i}th pulse and $\textbf{x}_i$ is the classical position of the atom \cite{Kasevich1992, Berman1997}.  For example, tilting $\textbf{k}_3$ by an additional angle $\delta \theta$ about the $x$-axis yields a phase $\Phi_\horizontal = k_\text{eff} \, \delta\theta \, y_3$ across the cloud, where $y_3$ is the horizontal position at the third pulse [Fig.~\ref{Fig:Apparatus}(a)].  This phase shear is independent of the details of the previous atom-laser interactions and of the implementation of the atomic source (in particular, its spatial extent, temperature, and quantum degeneracy).

Figure~\ref{Fig:Apparatus}(b) shows an image of the interferometer output that results from this horizontal phase shear, with $\delta \theta = 60~\mu\text{rad}$.  An optical ``pushing'' pulse, $5~\mu\text{s}$ long and resonant with the $\left|F=2\right\rangle \rightarrow \left|F'=3\right\rangle$ transition, separates the interferometer output ports.  Complementary fringes appear across each port, corresponding to the spatial variation of the atomic transition probability that results from phase shear.  For linear shears, the atom distribution at each port is modulated by an interference term $P(\textbf{r}) = \frac{1}{2} + \frac{C}{2} \sin(\boldsymbol\kappa\cdot\textbf{r} + \phi_0)$, where $C$ is the contrast, $\phi_0$ is the overall interferometer phase, and $\boldsymbol\kappa$ is the wavevector of the spatially varying component of the phase.

Since the retroreflection mirror can be tilted about an arbitrary horizontal axis, beam-tilt PSR can yield fringe patterns with $\hat{\boldsymbol{\kappa}}$ anywhere in the $xy$ plane, orthogonal to the laser beam axis [see Fig.~\ref{Fig:Apparatus}(a)].  For instance, it is possible to choose a tilt axis parallel to the line-of-sight of either of the CCD cameras (which are perpendicular), in which case we see a spatial fringe pattern with one camera, but no contrast with the other.  Hereafter, we tilt about the $x$-axis, yielding fringes on CCD2.

\begin{figure}
\begin{center}
\includegraphics[width=\columnwidth]{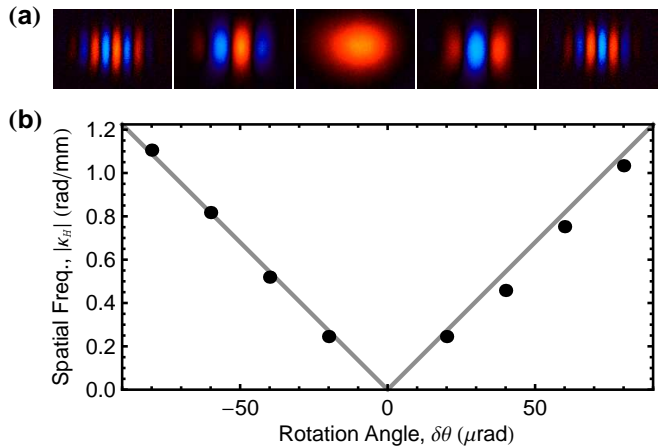}
\caption{ \label{Fig:fringeControl} Horizontal fringes resulting from beam-tilt PSR in a $2\,T=2.3~\text{s}$ interferometer.  (a) Spatial fringes observed on CCD2 with third-pulse tilt angles $\delta \theta = -80, -40, 0, +40, +80\,\mu\text{rad}$ (from left to right).  Red versus blue regions show anti-correlation in atom population.  Each image is the second-highest variance principal component arising from a set of 20 fluorescence images \cite{Dickerson2013}. (b) Measured fringe spatial frequency $|\kappa_\horizontal|$, resulting from images filtered using principal component analysis \cite{Dickerson2013}. We bin the images vertically and fit a Gaussian modulated by the interference term $P(\textbf{r})$.  The curve is a theoretical prediction with no free parameters.}
\end{center}
\end{figure}

The spatial frequency $\kappa$ of beam-tilt PSR fringes is set by the tilt angle $\delta \theta$. Figure~\ref{Fig:fringeControl}(b) shows the expected linear dependence, and it is apparent that by appropriate choice of the shear angle, the period of the shear can be tuned to an arbitrary value. While high spatial frequencies are desirable, in practice spatial frequency is limited by the depth of focus of the imaging system. Because we detect the atoms at a final drift time $t_d = 2.7~\text{s}$ that is later than the third pulse time $t_3 = 2.5~\text{s}$ (both measured from the time of trap release), we must correct for the continued motion of the atoms. In the limit where the initial size of the atomic source is much less than the final spatial extent of the atomic cloud (point source limit \cite{Dickerson2013,imagingArtifact}), the position at $t_d$ of an atom with velocity $v_y$ is $y\approx v_y t_d\approx y_3 \, t_d/t_3$. The detected horizontal fringe spatial frequency is then $\kappa_\horizontal \equiv \partial_y\Phi_\horizontal = k_\text{eff} \, \delta\theta \, t_3/t_d$.

\begin{figure}
\begin{center}
\includegraphics[width=\columnwidth]{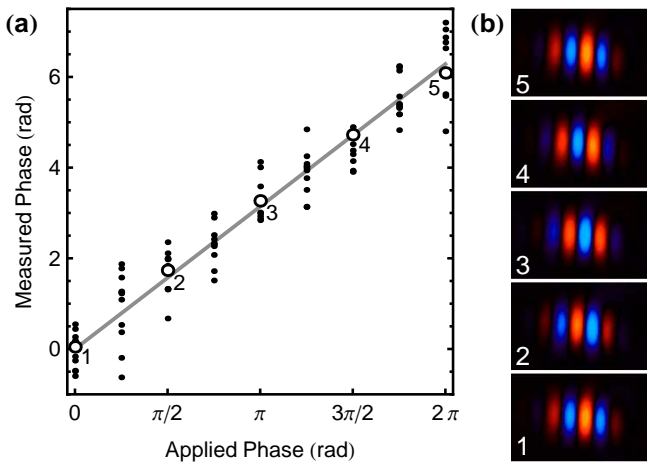}
\caption{ \label{Fig:PSRShortTLateTimePhaseScan} Demonstration of single-shot phase readout with a $2 \, T = 50~\text{ms}$ interferometer. (a) Measured phase versus the applied phase of the final atom-optics pulse for 96 shots. A line with unity slope is shown for reference. The measured phase is fit from images like those in (b). The measurement scatter at each phase step is dominated by technical noise introduced by vibration of the Raman laser beam delivery optics. (b) Five sample interferometer shots [open circles in (a)], separated in measured phase by $\sim\pi/2~\text{rad}$.  All images are filtered with principal component analysis.}
\end{center}
\end{figure}

To demonstrate single-shot phase readout, we implement a short interferometer sequence ($2 \, T = 50~\text{ms}$) near the end of the drift time. In this case, the atom cloud has a large spatial extent for the entire pulse sequence.  For each shot, we set the interferometer phase with an acousto-optic modulator and read it back using beam-tilt PSR with $\delta\theta = 60~\mu\text{rad}$.  Figure~\ref{Fig:PSRShortTLateTimePhaseScan} shows the expected correspondence between the applied and measured phases.  The spread in the measured phase is due to technical noise associated with spurious vibrations of the optics for the laser beams that drive the stimulated Raman transitions.

As an example of how PSR can enable a precision measurement, we implement an atom interferometric gyrocompass in a long interrogation time ($2T = 2.3~\text{s}$) configuration.  In this case, the Raman laser axis is rotated to compensate Earth's rotation, keeping this axis inertially fixed throughout the interrogation sequence. At the latitude of our lab in Stanford, California, this corresponds to an effective rotation rate of $\Omega_\earth = 57.9~\mu\text{rad}/\text{s}$ about an axis along the local true North vector, which we take to be at angle $\phi_\earth$ with respect to the $x$-axis. However, a small misalignment $\delta \phi_\earth \ll 1$ between the rotation axis of the retroreflection mirror and true North results in a residual rotation $\v{\delta\Omega}\approx\delta\phi_\earth\Omega_\earth \left(\sin{\phi_\earth}\v{\hat{x}}-\cos{\phi_\earth}\v{\hat{y}}\right)$ that leads to a Coriolis phase shift $\Phi_\coriolis = 2 \v{k_\text{eff}} \cdot \left(\v{\delta\Omega}\times\v{v}\right)T^2$ that varies across the cloud. As before, in the point source limit $v_y\approx y/t_d$, so the Coriolis phase gradient is $\kappa_{\coriolis,y}\equiv\partial_y\Phi_\coriolis = 2 k_\text{eff} T^2 \delta\phi_\earth\Omega_\earth \sin{\phi_\earth}/t_d$.  To realize a gyrocompass, we vary the axis of applied rotation by scanning $\delta \phi_\earth$, and identify true North with the angle at which $\kappa_{\coriolis,y}=0$.

It can be challenging to measure small phase gradients with spatial frequencies $\kappa \ll 1/\sigma$, where $\sigma$ is the width of the atom ensemble. In this limit, there is much less than one fringe period across the cloud, so the fringe fitting method shown in Fig.~\ref{Fig:fringeControl}(b) cannot be used. Instead, the gradient can be estimated by measuring phase differences across the ensemble (e.g., with ellipse fits \cite{Foster2002}), but this procedure can be sensitive to fluctuations in the atomic density distribution (width, position, and shape).

To circumvent these issues, we take advantage of PSR by applying an additional phase shear that augments the residual Coriolis shear $\Phi_\coriolis$. An additional tilt of $\delta\theta=\pm60~\mu\text{rad}$ about the $x$-axis is added before the final interferometer pulse. This introduces a horizontal shear $\Phi_\horizontal$ with approximately $2.5$ fringe periods across the cloud, visible on CCD2. Depending on the sign of the tilt angle, this shear adds to or subtracts from $\Phi_\coriolis$. The combined phase gradient is then $\kappa_\pm\equiv  k_\text{eff}\left|\delta\theta\right| t_3/t_d  \,\pm\, \kappa_{\coriolis,y}$ and is large enough to use fringe fitting to extract the spatial frequency. This technique of shifting a small phase gradient to a larger spatial frequency is analogous to a heterodyne measurement in the time domain.  In both cases, the heterodyne process circumvents low frequency noise. By alternating the sign of the additional $60~\mu\text{rad}$ tilt, a differential measurement is possible whereby systematic uncertainty in the applied shear angle is mitigated: $\Delta\kappa\equiv\kappa_{+}-\kappa_{-}=2 \kappa_{\coriolis,y}$, independent of the magnitude of $\delta\theta$.

\begin{figure}[t]
\begin{center}
\includegraphics[width=\columnwidth]{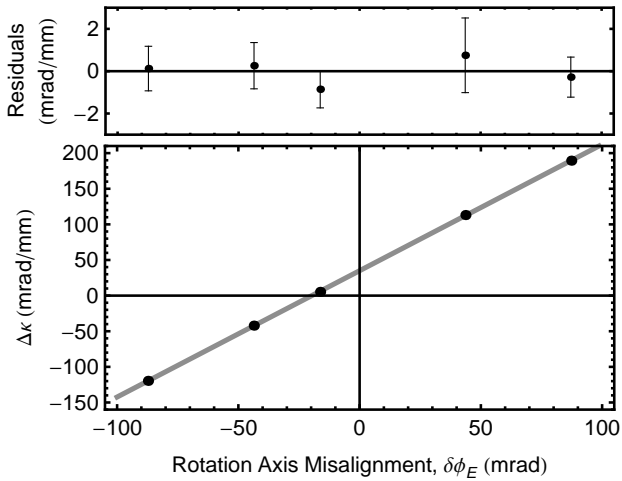}
\caption{ \label{Fig:GyrocompassingWithPSR} Gyrocompass using the phase shear method. Each $\Delta\kappa$ point is the combination of 40 trials, 20 at each of the two applied tilt values ($\delta\theta=\pm 60~\mu\text{rad}$). The horizontal intercept of a linear fit gives the direction of true North.}
\end{center}
\end{figure}

Figure~\ref{Fig:GyrocompassingWithPSR} shows the expected linear scaling of the differential spatial frequency $\Delta\kappa$ as a function of the applied rotation angle $\delta \phi_\earth$. A linear fit to the data yields a horizontal intercept that indicates the direction of true North with a precision of $10~\text{millidegrees}$. We note that an apparatus optimized for gyrocompass performance could achieve similar or better precision in a more compact form factor. Also, this method does not require a vibrationally stable environment since the measurement rests on the determination of the fringe period, not the overall phase.

\begin{figure}
\begin{center}
\includegraphics[width=\columnwidth]{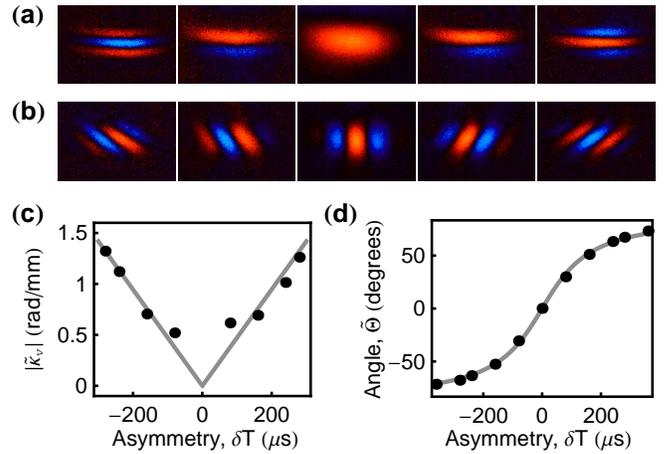}
\caption{ \label{Fig:arctan} Arbitrary control of spatial fringe direction. (a) Second-highest variance principal components from sets of 20 trials with timing asymmetry $\delta T = -240, -160, 0, +160, +240 \, \mu \text{s}$ (from left to right) (b) Comparable images for trials with both a beam tilt $\delta \theta = 40 \, \mu \text{rad}$ and $\delta T = -160, -80, 0, +80, +160 \, \mu \text{s}$. (c) Measured fringe spatial frequency extracted from fits to principal component filtered images with vertical fringes. (d) Measured fringe angle extracted from fits to images with tilted fringes. In both (c) and (d) the curves are predictions with no free parameters.}
\end{center}
\end{figure}

Finally, we show how combining beam tilts and interferometer timing asymmetries provides nearly arbitrary control over the spatial wavevector $\boldsymbol\kappa$ of the applied shear. While a beam tilt applies a phase shear with spatial wavevector in the plane transverse to the interferometer beam axis, interferometer timing asymmetry yields a phase shear parallel to the beam axis ($\boldsymbol\kappa \parallel \mathbf{k}_\text{eff}$) in the point source limit \cite{Muntinga2013}.  To create an asymmetric interferometer, we offset the central $\pi$ pulse by $\delta T/2$ such that the time between the first and second pulses ($T + \delta T/2$) is different from the time between the second and third pulses ($T - \delta T/2$). The resulting phase shift, $\Phi_\vertical = k_\text{eff} \, v_z \delta T$, depends on the atoms' Doppler shift along the direction of $\textbf{k}_\text{eff}$. The phase shear at detection is then $\kappa_\vertical = \partial_z \Phi_\vertical =  k_\text{eff} \, \delta T/ t_d$. Figure~\ref{Fig:arctan}(a) shows the resulting vertical fringes, which are orthogonal to those from beam tilts seen in Fig.~\ref{Fig:fringeControl}(a) and are simultaneously visible on both CCD cameras.  The fitted fringe frequency shown in Fig.~\ref{Fig:arctan}(c) exhibits the expected linear dependence as a function of $\delta T$, deviating at low spatial frequency due to the difficulty of fitting a fringe with $\kappa \sim 1/\sigma$.

For these vertical fringes, we find that the imaging pulse reduces the detected spatial frequency by stretching the cloud vertically. We independently characterize this stretch by measuring the vertical fringe period as a function of imaging duration $\tau$ and then extrapolating to $\tau = 0$. The results indicate a fraction stretch rate of $\alpha = 0.12~\text{ms}^{-1}$. The modified prediction for the spatial frequency is $\widetilde{\kappa}_\vertical = \kappa_\vertical / \left(1 + \alpha \tau \right)$.  With the $\tau = 2 \, \text{ms}$ imaging time used, this agrees well with the measurements of Fig.~\ref{Fig:arctan}(c) with no free parameters.

By combining beam tilt shear $\kappa_\horizontal$ with timing asymmetry shear $\kappa_\vertical$, we can create spatial fringes at arbitrary angles. The composite phase shear is at an angle $\Theta = \arctan{\left(\kappa_\vertical /\kappa_\horizontal \right)}=\arctan{\left[\delta T/ \left(\delta \theta \, t_3 \right)\right]}$. Figures~\ref{Fig:arctan}(b) and (d) show the fringe images and extracted angles using a $\delta \theta = 40 \, \mu \text{rad}$ beam tilt combined with a range of timing asymmetries.  To find the angles, we apply Fourier and principal component filters and fit with a two-dimensional Gaussian envelope modulated by an interference term $P(\textbf{r})$. Because the vertical stretch imparted by the imaging beams modifies the measured angle, we again correct for image stretching during detection. The modified prediction, $\widetilde{\Theta} = \arccot\left[(1+ \alpha \tau) \cot \Theta \right]$, shows good agreement with the measured angles of Fig.~\ref{Fig:arctan}(d) with no free parameters.

We have demonstrated a precision gyrocompass with PSR, but with arbitrary control of the shear angle the method can be used to measure phase shifts and gradients from any origin.  For example, a vertical gravity gradient $T_{zz}$ induces a phase shear $k_\text{eff}\, T_{zz} v_z T^3$. This shear translates the measured angles of Fig.~\ref{Fig:arctan}(d) such that $\Theta = \arctan{\left[\left(\delta T - T_{zz}T^3\right)/\left(\delta \theta \, t_3\right)\right]}$.  For our parameters, this would yield an effective asymmetry of $2~\text{ns}/\text{E}$. PSR can also be used to measure nonlinear phase variations, including optical wavefront aberrations \cite{Dickerson2013}. Finally, we expect the phase shear method to be enabling for future inertial sensors operating on dynamic platforms, where single shot estimation of phase and contrast is vital.

\begin{acknowledgments}
The authors would like to thank Philippe~Bouyer, Sheng-wey~Chiow, Tim~Kovachy, and Jan~Rudolph for valuable discussions and contributions to the apparatus. AS acknowledges support from the NSF GRFP.  SMD acknowledges support from the Gerald J. Lieberman Fellowship. This work was supported in part by NASA GSFC Grant No. NNX11AM31A.
\end{acknowledgments}

\bibliographystyle{apsrev4-1}

\bibliography{EPKasevich-PSR}

\begin{thebibliography}{19}%
\makeatletter
\providecommand \@ifxundefined [1]{%
 \@ifx{#1\undefined}
}%
\providecommand \@ifnum [1]{%
 \ifnum #1\expandafter \@firstoftwo
 \else \expandafter \@secondoftwo
 \fi
}%
\providecommand \@ifx [1]{%
 \ifx #1\expandafter \@firstoftwo
 \else \expandafter \@secondoftwo
 \fi
}%
\providecommand \natexlab [1]{#1}%
\providecommand \enquote  [1]{``#1''}%
\providecommand \bibnamefont  [1]{#1}%
\providecommand \bibfnamefont [1]{#1}%
\providecommand \citenamefont [1]{#1}%
\providecommand \href@noop [0]{\@secondoftwo}%
\providecommand \href [0]{\begingroup \@sanitize@url \@href}%
\providecommand \@href[1]{\@@startlink{#1}\@@href}%
\providecommand \@@href[1]{\endgroup#1\@@endlink}%
\providecommand \@sanitize@url [0]{\catcode `\\12\catcode `\$12\catcode
  `\&12\catcode `\#12\catcode `\^12\catcode `\_12\catcode `\%12\relax}%
\providecommand \@@startlink[1]{}%
\providecommand \@@endlink[0]{}%
\providecommand \url  [0]{\begingroup\@sanitize@url \@url }%
\providecommand \@url [1]{\endgroup\@href {#1}{\urlprefix }}%
\providecommand \urlprefix  [0]{URL }%
\providecommand \Eprint [0]{\href }%
\providecommand \doibase [0]{http://dx.doi.org/}%
\providecommand \selectlanguage [0]{\@gobble}%
\providecommand \bibinfo  [0]{\@secondoftwo}%
\providecommand \bibfield  [0]{\@secondoftwo}%
\providecommand \translation [1]{[#1]}%
\providecommand \BibitemOpen [0]{}%
\providecommand \bibitemStop [0]{}%
\providecommand \bibitemNoStop [0]{.\EOS\space}%
\providecommand \EOS [0]{\spacefactor3000\relax}%
\providecommand \BibitemShut  [1]{\csname bibitem#1\endcsname}%
\let\auto@bib@innerbib\@empty
\bibitem [{\citenamefont {Berman}(1997)}]{Berman1997}%
  \BibitemOpen
  \bibinfo {editor} {\bibfnamefont {P.~R.}\ \bibnamefont {Berman}},\ ed.,\
  \href@noop {} {\emph {\bibinfo {title} {{Atom Interferometry}}}}\ (\bibinfo
  {publisher} {Academic Press},\ \bibinfo {address} {San Diego},\ \bibinfo
  {year} {1997})\BibitemShut {NoStop}%
\bibitem [{\citenamefont {Fixler}\ \emph {et~al.}(2007)\citenamefont {Fixler},
  \citenamefont {Foster}, \citenamefont {McGuirk},\ and\ \citenamefont
  {Kasevich}}]{Fixler2007}%
  \BibitemOpen
  \bibfield  {author} {\bibinfo {author} {\bibfnamefont {J.~B.}\ \bibnamefont
  {Fixler}}, \bibinfo {author} {\bibfnamefont {G.~T.}\ \bibnamefont {Foster}},
  \bibinfo {author} {\bibfnamefont {J.~M.}\ \bibnamefont {McGuirk}}, \ and\
  \bibinfo {author} {\bibfnamefont {M.~A.}\ \bibnamefont {Kasevich}},\ }\href
  {\doibase 10.1126/science.1135459} {\bibfield  {journal} {\bibinfo  {journal}
  {Science}\ }\textbf {\bibinfo {volume} {315}},\ \bibinfo {pages} {74}
  (\bibinfo {year} {2007})}\BibitemShut {NoStop}%
\bibitem [{\citenamefont {Lamporesi}\ \emph {et~al.}(2008)\citenamefont
  {Lamporesi}, \citenamefont {Bertoldi}, \citenamefont {Cacciapuoti},
  \citenamefont {Prevedelli},\ and\ \citenamefont {Tino}}]{Lamporesi2008}%
  \BibitemOpen
  \bibfield  {author} {\bibinfo {author} {\bibfnamefont {G.}~\bibnamefont
  {Lamporesi}}, \bibinfo {author} {\bibfnamefont {A.}~\bibnamefont {Bertoldi}},
  \bibinfo {author} {\bibfnamefont {L.}~\bibnamefont {Cacciapuoti}}, \bibinfo
  {author} {\bibfnamefont {M.}~\bibnamefont {Prevedelli}}, \ and\ \bibinfo
  {author} {\bibfnamefont {G.}~\bibnamefont {Tino}},\ }\href {\doibase
  10.1103/PhysRevLett.100.050801} {\bibfield  {journal} {\bibinfo  {journal}
  {Physical Review Letters}\ }\textbf {\bibinfo {volume} {100}},\ \bibinfo
  {pages} {050801} (\bibinfo {year} {2008})}\BibitemShut {NoStop}%
\bibitem [{\citenamefont {Bouchendira}\ \emph {et~al.}(2011)\citenamefont
  {Bouchendira}, \citenamefont {Clad\'{e}}, \citenamefont
  {Guellati-Kh\'{e}lifa}, \citenamefont {Nez},\ and\ \citenamefont
  {Biraben}}]{Bouchendira2011}%
  \BibitemOpen
  \bibfield  {author} {\bibinfo {author} {\bibfnamefont {R.}~\bibnamefont
  {Bouchendira}}, \bibinfo {author} {\bibfnamefont {P.}~\bibnamefont
  {Clad\'{e}}}, \bibinfo {author} {\bibfnamefont {S.}~\bibnamefont
  {Guellati-Kh\'{e}lifa}}, \bibinfo {author} {\bibfnamefont {F.}~\bibnamefont
  {Nez}}, \ and\ \bibinfo {author} {\bibfnamefont {F.}~\bibnamefont
  {Biraben}},\ }\href {\doibase 10.1103/PhysRevLett.106.080801} {\bibfield
  {journal} {\bibinfo  {journal} {Physical Review Letters}\ }\textbf {\bibinfo
  {volume} {106}},\ \bibinfo {pages} {080801} (\bibinfo {year}
  {2011})}\BibitemShut {NoStop}%
\bibitem [{\citenamefont {Hogan}\ \emph {et~al.}(2009)\citenamefont {Hogan},
  \citenamefont {Johnson},\ and\ \citenamefont {Kasevich}}]{Hogan2009}%
  \BibitemOpen
  \bibfield  {author} {\bibinfo {author} {\bibfnamefont {J.~M.}\ \bibnamefont
  {Hogan}}, \bibinfo {author} {\bibfnamefont {D.~M.~S.}\ \bibnamefont
  {Johnson}}, \ and\ \bibinfo {author} {\bibfnamefont {M.~A.}\ \bibnamefont
  {Kasevich}},\ }in\ \href {http://arxiv.org/abs/0806.3261} {\emph {\bibinfo
  {booktitle} {Proceedings of the International School of Physics "Enrico
  Fermi" on Atom Optics and Space Physics}}},\ \bibinfo {editor} {edited by\
  \bibinfo {editor} {\bibfnamefont {E.}~\bibnamefont {Arimondo}}, \bibinfo
  {editor} {\bibfnamefont {W.}~\bibnamefont {Ertmer}}, \ and\ \bibinfo {editor}
  {\bibfnamefont {W.~P.}\ \bibnamefont {Schleich}}}\ (\bibinfo  {publisher}
  {IOS Press},\ \bibinfo {address} {Amsterdam},\ \bibinfo {year} {2009})\ pp.\
  \bibinfo {pages} {411--447},\ \Eprint {http://arxiv.org/abs/0806.3261}
  {arXiv:0806.3261} \BibitemShut {NoStop}%
\bibitem [{\citenamefont {Dimopoulos}\ \emph {et~al.}(2007)\citenamefont
  {Dimopoulos}, \citenamefont {Graham}, \citenamefont {Hogan},\ and\
  \citenamefont {Kasevich}}]{Dimopoulos2007}%
  \BibitemOpen
  \bibfield  {author} {\bibinfo {author} {\bibfnamefont {S.}~\bibnamefont
  {Dimopoulos}}, \bibinfo {author} {\bibfnamefont {P.~W.}\ \bibnamefont
  {Graham}}, \bibinfo {author} {\bibfnamefont {J.~M.}\ \bibnamefont {Hogan}}, \
  and\ \bibinfo {author} {\bibfnamefont {M.~A.}\ \bibnamefont {Kasevich}},\
  }\href {\doibase 10.1103/PhysRevLett.98.111102} {\bibfield  {journal}
  {\bibinfo  {journal} {Physical Review Letters}\ }\textbf {\bibinfo {volume}
  {98}},\ \bibinfo {pages} {111102} (\bibinfo {year} {2007})}\BibitemShut
  {NoStop}%
\bibitem [{\citenamefont {Hohensee}\ \emph {et~al.}(2011)\citenamefont
  {Hohensee}, \citenamefont {Chu}, \citenamefont {Peters},\ and\ \citenamefont
  {M\"{u}ller}}]{Hohensee2011}%
  \BibitemOpen
  \bibfield  {author} {\bibinfo {author} {\bibfnamefont {M.~A.}\ \bibnamefont
  {Hohensee}}, \bibinfo {author} {\bibfnamefont {S.}~\bibnamefont {Chu}},
  \bibinfo {author} {\bibfnamefont {A.}~\bibnamefont {Peters}}, \ and\ \bibinfo
  {author} {\bibfnamefont {H.}~\bibnamefont {M\"{u}ller}},\ }\href {\doibase
  10.1103/PhysRevLett.106.151102} {\bibfield  {journal} {\bibinfo  {journal}
  {Physical Review Letters}\ }\textbf {\bibinfo {volume} {106}},\ \bibinfo
  {pages} {151102} (\bibinfo {year} {2011})}\BibitemShut {NoStop}%
\bibitem [{\citenamefont {Dimopoulos}\ \emph {et~al.}(2008)\citenamefont
  {Dimopoulos}, \citenamefont {Graham}, \citenamefont {Hogan}, \citenamefont
  {Kasevich},\ and\ \citenamefont {Rajendran}}]{Dimopoulos2008b}%
  \BibitemOpen
  \bibfield  {author} {\bibinfo {author} {\bibfnamefont {S.}~\bibnamefont
  {Dimopoulos}}, \bibinfo {author} {\bibfnamefont {P.~W.}\ \bibnamefont
  {Graham}}, \bibinfo {author} {\bibfnamefont {J.~M.}\ \bibnamefont {Hogan}},
  \bibinfo {author} {\bibfnamefont {M.~A.}\ \bibnamefont {Kasevich}}, \ and\
  \bibinfo {author} {\bibfnamefont {S.}~\bibnamefont {Rajendran}},\ }\href
  {\doibase 10.1103/PhysRevD.78.122002} {\bibfield  {journal} {\bibinfo
  {journal} {Physical Review D}\ }\textbf {\bibinfo {volume} {78}},\ \bibinfo
  {pages} {122002} (\bibinfo {year} {2008})}\BibitemShut {NoStop}%
\bibitem [{\citenamefont {Graham}\ \emph {et~al.}(2013)\citenamefont {Graham},
  \citenamefont {Hogan}, \citenamefont {Kasevich},\ and\ \citenamefont
  {Rajendran}}]{Graham2013}%
  \BibitemOpen
  \bibfield  {author} {\bibinfo {author} {\bibfnamefont {P.~W.}\ \bibnamefont
  {Graham}}, \bibinfo {author} {\bibfnamefont {J.~M.}\ \bibnamefont {Hogan}},
  \bibinfo {author} {\bibfnamefont {M.~A.}\ \bibnamefont {Kasevich}}, \ and\
  \bibinfo {author} {\bibfnamefont {S.}~\bibnamefont {Rajendran}},\ }\href
  {\doibase 10.1103/PhysRevLett.110.171102} {\bibfield  {journal} {\bibinfo
  {journal} {Physical Review Letters}\ }\textbf {\bibinfo {volume} {110}},\
  \bibinfo {pages} {171102} (\bibinfo {year} {2013})}\BibitemShut {NoStop}%
\bibitem [{\citenamefont {Gustavson}\ \emph {et~al.}(1997)\citenamefont
  {Gustavson}, \citenamefont {Bouyer},\ and\ \citenamefont
  {Kasevich}}]{Gustavson1997}%
  \BibitemOpen
  \bibfield  {author} {\bibinfo {author} {\bibfnamefont {T.~L.}\ \bibnamefont
  {Gustavson}}, \bibinfo {author} {\bibfnamefont {P.}~\bibnamefont {Bouyer}}, \
  and\ \bibinfo {author} {\bibfnamefont {M.~A.}\ \bibnamefont {Kasevich}},\
  }\href {\doibase 10.1103/PhysRevLett.78.2046} {\bibfield  {journal} {\bibinfo
   {journal} {Physical Review Letters}\ }\textbf {\bibinfo {volume} {78}},\
  \bibinfo {pages} {2046} (\bibinfo {year} {1997})}\BibitemShut {NoStop}%
\bibitem [{\citenamefont {Geiger}\ \emph {et~al.}(2011)\citenamefont {Geiger},
  \citenamefont {M\'{e}noret}, \citenamefont {Stern}, \citenamefont {Zahzam},
  \citenamefont {Cheinet}, \citenamefont {Battelier}, \citenamefont {Villing},
  \citenamefont {Moron}, \citenamefont {Lours}, \citenamefont {Bidel},
  \citenamefont {Bresson}, \citenamefont {Landragin},\ and\ \citenamefont
  {Bouyer}}]{Geiger2011}%
  \BibitemOpen
  \bibfield  {author} {\bibinfo {author} {\bibfnamefont {R.}~\bibnamefont
  {Geiger}}, \bibinfo {author} {\bibfnamefont {V.}~\bibnamefont {M\'{e}noret}},
  \bibinfo {author} {\bibfnamefont {G.}~\bibnamefont {Stern}}, \bibinfo
  {author} {\bibfnamefont {N.}~\bibnamefont {Zahzam}}, \bibinfo {author}
  {\bibfnamefont {P.}~\bibnamefont {Cheinet}}, \bibinfo {author} {\bibfnamefont
  {B.}~\bibnamefont {Battelier}}, \bibinfo {author} {\bibfnamefont
  {A.}~\bibnamefont {Villing}}, \bibinfo {author} {\bibfnamefont
  {F.}~\bibnamefont {Moron}}, \bibinfo {author} {\bibfnamefont
  {M.}~\bibnamefont {Lours}}, \bibinfo {author} {\bibfnamefont
  {Y.}~\bibnamefont {Bidel}}, \bibinfo {author} {\bibfnamefont
  {A.}~\bibnamefont {Bresson}}, \bibinfo {author} {\bibfnamefont
  {A.}~\bibnamefont {Landragin}}, \ and\ \bibinfo {author} {\bibfnamefont
  {P.}~\bibnamefont {Bouyer}},\ }\href {\doibase 10.1038/ncomms1479} {\bibfield
   {journal} {\bibinfo  {journal} {Nature Communications}\ }\textbf {\bibinfo
  {volume} {2}},\ \bibinfo {pages} {474} (\bibinfo {year} {2011})}\BibitemShut
  {NoStop}%
\bibitem [{\citenamefont {Peters}\ \emph {et~al.}(2001)\citenamefont {Peters},
  \citenamefont {Chung},\ and\ \citenamefont {Chu}}]{Peters2001}%
  \BibitemOpen
  \bibfield  {author} {\bibinfo {author} {\bibfnamefont {A.}~\bibnamefont
  {Peters}}, \bibinfo {author} {\bibfnamefont {K.~Y.}\ \bibnamefont {Chung}}, \
  and\ \bibinfo {author} {\bibfnamefont {S.}~\bibnamefont {Chu}},\ }\href
  {\doibase 10.1088/0026-1394/38/1/4} {\bibfield  {journal} {\bibinfo
  {journal} {Metrologia}\ }\textbf {\bibinfo {volume} {38}},\ \bibinfo {pages}
  {25} (\bibinfo {year} {2001})}\BibitemShut {NoStop}%
\bibitem [{\citenamefont {McGuirk}\ \emph {et~al.}(2002)\citenamefont
  {McGuirk}, \citenamefont {Foster}, \citenamefont {Fixler}, \citenamefont
  {Snadden},\ and\ \citenamefont {Kasevich}}]{McGuirk2002}%
  \BibitemOpen
  \bibfield  {author} {\bibinfo {author} {\bibfnamefont {J.~M.}\ \bibnamefont
  {McGuirk}}, \bibinfo {author} {\bibfnamefont {G.~T.}\ \bibnamefont {Foster}},
  \bibinfo {author} {\bibfnamefont {J.~B.}\ \bibnamefont {Fixler}}, \bibinfo
  {author} {\bibfnamefont {M.~J.}\ \bibnamefont {Snadden}}, \ and\ \bibinfo
  {author} {\bibfnamefont {M.~A.}\ \bibnamefont {Kasevich}},\ }\href {\doibase
  10.1103/PhysRevA.65.033608} {\bibfield  {journal} {\bibinfo  {journal}
  {Physical Review A}\ }\textbf {\bibinfo {volume} {65}},\ \bibinfo {pages}
  {033608} (\bibinfo {year} {2002})}\BibitemShut {NoStop}%
\bibitem [{\citenamefont {Kasevich}\ and\ \citenamefont
  {Chu}(1992)}]{Kasevich1992}%
  \BibitemOpen
  \bibfield  {author} {\bibinfo {author} {\bibfnamefont {M.}~\bibnamefont
  {Kasevich}}\ and\ \bibinfo {author} {\bibfnamefont {S.}~\bibnamefont {Chu}},\
  }\href {\doibase 10.1007/BF00325375} {\bibfield  {journal} {\bibinfo
  {journal} {Applied Physics B Photophysics and Laser Chemistry}\ }\textbf
  {\bibinfo {volume} {54}},\ \bibinfo {pages} {321} (\bibinfo {year}
  {1992})}\BibitemShut {NoStop}%
\bibitem [{\citenamefont {Dickerson}\ \emph {et~al.}(2013)\citenamefont
  {Dickerson}, \citenamefont {Hogan}, \citenamefont {Sugarbaker}, \citenamefont
  {Johnson},\ and\ \citenamefont {Kasevich}}]{Dickerson2013}%
  \BibitemOpen
  \bibfield  {author} {\bibinfo {author} {\bibfnamefont {S.~M.}\ \bibnamefont
  {Dickerson}}, \bibinfo {author} {\bibfnamefont {J.~M.}\ \bibnamefont
  {Hogan}}, \bibinfo {author} {\bibfnamefont {A.}~\bibnamefont {Sugarbaker}},
  \bibinfo {author} {\bibfnamefont {D.~M.~S.}\ \bibnamefont {Johnson}}, \ and\
  \bibinfo {author} {\bibfnamefont {M.~A.}\ \bibnamefont {Kasevich}},\
  }\href@noop {} {\  (\bibinfo {year} {2013})},\ \Eprint
  {http://arxiv.org/abs/arXiv:1305.1700} {arXiv:1305.1700} \BibitemShut
  {NoStop}%
\bibitem [{\citenamefont {Dickerson}\ \emph {et~al.}(2012)\citenamefont
  {Dickerson}, \citenamefont {Hogan}, \citenamefont {Johnson}, \citenamefont
  {Kovachy}, \citenamefont {Sugarbaker}, \citenamefont {Chiow},\ and\
  \citenamefont {Kasevich}}]{Dickerson2012}%
  \BibitemOpen
  \bibfield  {author} {\bibinfo {author} {\bibfnamefont {S.}~\bibnamefont
  {Dickerson}}, \bibinfo {author} {\bibfnamefont {J.~M.}\ \bibnamefont
  {Hogan}}, \bibinfo {author} {\bibfnamefont {D.~M.~S.}\ \bibnamefont
  {Johnson}}, \bibinfo {author} {\bibfnamefont {T.}~\bibnamefont {Kovachy}},
  \bibinfo {author} {\bibfnamefont {A.}~\bibnamefont {Sugarbaker}}, \bibinfo
  {author} {\bibfnamefont {S.-w.}\ \bibnamefont {Chiow}}, \ and\ \bibinfo
  {author} {\bibfnamefont {M.~A.}\ \bibnamefont {Kasevich}},\ }\href {\doibase
  10.1063/1.4720943} {\bibfield  {journal} {\bibinfo  {journal} {The Review of
  Scientific Instruments}\ }\textbf {\bibinfo {volume} {83}},\ \bibinfo {pages}
  {065108} (\bibinfo {year} {2012})}\BibitemShut {NoStop}%
\bibitem [{ima()}]{imagingArtifact}%
  \BibitemOpen
  \href@noop {} {}\bibinfo {note} {The point source limit is relevant here only
  as an imaging artifact, resulting from $t_d > t_3$. The point source limit is
  not necessary for beam-tilt PSR.}\BibitemShut {Stop}%
\bibitem [{\citenamefont {Foster}\ \emph {et~al.}(2002)\citenamefont {Foster},
  \citenamefont {Fixler}, \citenamefont {McGuirk},\ and\ \citenamefont
  {Kasevich}}]{Foster2002}%
  \BibitemOpen
  \bibfield  {author} {\bibinfo {author} {\bibfnamefont {G.~T.}\ \bibnamefont
  {Foster}}, \bibinfo {author} {\bibfnamefont {J.~B.}\ \bibnamefont {Fixler}},
  \bibinfo {author} {\bibfnamefont {J.~M.}\ \bibnamefont {McGuirk}}, \ and\
  \bibinfo {author} {\bibfnamefont {M.~A.}\ \bibnamefont {Kasevich}},\ }\href
  {http://www.opticsinfobase.org/abstract.cfm?\&id=68960} {\bibfield  {journal}
  {\bibinfo  {journal} {Optics Letters}\ }\textbf {\bibinfo {volume} {27}},\
  \bibinfo {pages} {951} (\bibinfo {year} {2002})}\BibitemShut {NoStop}%
\bibitem [{\citenamefont {M\"{u}ntinga}\ \emph {et~al.}(2013)\citenamefont
  {M\"{u}ntinga}, \citenamefont {Ahlers}, \citenamefont {Krutzik},
  \citenamefont {Wenzlawski}, \citenamefont {Arnold}, \citenamefont {Becker},
  \citenamefont {Bongs}, \citenamefont {Dittus}, \citenamefont {Duncker},
  \citenamefont {Gaaloul}, \citenamefont {Gherasim}, \citenamefont {Giese},
  \citenamefont {Grzeschik}, \citenamefont {H\"{a}nsch}, \citenamefont
  {Hellmig}, \citenamefont {Herr}, \citenamefont {Herrmann}, \citenamefont
  {Kajari}, \citenamefont {Kleinert}, \citenamefont {L\"{a}mmerzahl},
  \citenamefont {Lewoczko-Adamczyk}, \citenamefont {Malcolm}, \citenamefont
  {Meyer}, \citenamefont {Nolte}, \citenamefont {Peters}, \citenamefont {Popp},
  \citenamefont {Reichel}, \citenamefont {Roura}, \citenamefont {Rudolph},
  \citenamefont {Schiemangk}, \citenamefont {Schneider}, \citenamefont
  {Seidel}, \citenamefont {Sengstock}, \citenamefont {Tamma}, \citenamefont
  {Valenzuela}, \citenamefont {Vogel}, \citenamefont {Walser}, \citenamefont
  {Wendrich}, \citenamefont {Windpassinger}, \citenamefont {Zeller},
  \citenamefont {van Zoest}, \citenamefont {Ertmer}, \citenamefont {Schleich},\
  and\ \citenamefont {Rasel}}]{Muntinga2013}%
  \BibitemOpen
  \bibfield  {author} {\bibinfo {author} {\bibfnamefont {H.}~\bibnamefont
  {M\"{u}ntinga}}, \bibinfo {author} {\bibfnamefont {H.}~\bibnamefont
  {Ahlers}}, \bibinfo {author} {\bibfnamefont {M.}~\bibnamefont {Krutzik}},
  \bibinfo {author} {\bibfnamefont {A.}~\bibnamefont {Wenzlawski}}, \bibinfo
  {author} {\bibfnamefont {S.}~\bibnamefont {Arnold}}, \bibinfo {author}
  {\bibfnamefont {D.}~\bibnamefont {Becker}}, \bibinfo {author} {\bibfnamefont
  {K.}~\bibnamefont {Bongs}}, \bibinfo {author} {\bibfnamefont
  {H.}~\bibnamefont {Dittus}}, \bibinfo {author} {\bibfnamefont
  {H.}~\bibnamefont {Duncker}}, \bibinfo {author} {\bibfnamefont
  {N.}~\bibnamefont {Gaaloul}}, \bibinfo {author} {\bibfnamefont
  {C.}~\bibnamefont {Gherasim}}, \bibinfo {author} {\bibfnamefont
  {E.}~\bibnamefont {Giese}}, \bibinfo {author} {\bibfnamefont
  {C.}~\bibnamefont {Grzeschik}}, \bibinfo {author} {\bibfnamefont {T.~W.}\
  \bibnamefont {H\"{a}nsch}}, \bibinfo {author} {\bibfnamefont
  {O.}~\bibnamefont {Hellmig}}, \bibinfo {author} {\bibfnamefont
  {W.}~\bibnamefont {Herr}}, \bibinfo {author} {\bibfnamefont {S.}~\bibnamefont
  {Herrmann}}, \bibinfo {author} {\bibfnamefont {E.}~\bibnamefont {Kajari}},
  \bibinfo {author} {\bibfnamefont {S.}~\bibnamefont {Kleinert}}, \bibinfo
  {author} {\bibfnamefont {C.}~\bibnamefont {L\"{a}mmerzahl}}, \bibinfo
  {author} {\bibfnamefont {W.}~\bibnamefont {Lewoczko-Adamczyk}}, \bibinfo
  {author} {\bibfnamefont {J.}~\bibnamefont {Malcolm}}, \bibinfo {author}
  {\bibfnamefont {N.}~\bibnamefont {Meyer}}, \bibinfo {author} {\bibfnamefont
  {R.}~\bibnamefont {Nolte}}, \bibinfo {author} {\bibfnamefont
  {A.}~\bibnamefont {Peters}}, \bibinfo {author} {\bibfnamefont
  {M.}~\bibnamefont {Popp}}, \bibinfo {author} {\bibfnamefont {J.}~\bibnamefont
  {Reichel}}, \bibinfo {author} {\bibfnamefont {A.}~\bibnamefont {Roura}},
  \bibinfo {author} {\bibfnamefont {J.}~\bibnamefont {Rudolph}}, \bibinfo
  {author} {\bibfnamefont {M.}~\bibnamefont {Schiemangk}}, \bibinfo {author}
  {\bibfnamefont {M.}~\bibnamefont {Schneider}}, \bibinfo {author}
  {\bibfnamefont {S.~T.}\ \bibnamefont {Seidel}}, \bibinfo {author}
  {\bibfnamefont {K.}~\bibnamefont {Sengstock}}, \bibinfo {author}
  {\bibfnamefont {V.}~\bibnamefont {Tamma}}, \bibinfo {author} {\bibfnamefont
  {T.}~\bibnamefont {Valenzuela}}, \bibinfo {author} {\bibfnamefont
  {A.}~\bibnamefont {Vogel}}, \bibinfo {author} {\bibfnamefont
  {R.}~\bibnamefont {Walser}}, \bibinfo {author} {\bibfnamefont
  {T.}~\bibnamefont {Wendrich}}, \bibinfo {author} {\bibfnamefont
  {P.}~\bibnamefont {Windpassinger}}, \bibinfo {author} {\bibfnamefont
  {W.}~\bibnamefont {Zeller}}, \bibinfo {author} {\bibfnamefont
  {T.}~\bibnamefont {van Zoest}}, \bibinfo {author} {\bibfnamefont
  {W.}~\bibnamefont {Ertmer}}, \bibinfo {author} {\bibfnamefont {W.~P.}\
  \bibnamefont {Schleich}}, \ and\ \bibinfo {author} {\bibfnamefont {E.~M.}\
  \bibnamefont {Rasel}},\ }\href {\doibase 10.1103/PhysRevLett.110.093602}
  {\bibfield  {journal} {\bibinfo  {journal} {Physical Review Letters}\
  }\textbf {\bibinfo {volume} {110}},\ \bibinfo {pages} {093602} (\bibinfo
  {year} {2013})}\BibitemShut {NoStop}%
\end{thebibliography}%

\end{document}